\documentclass[conference]{IEEEtran}
\IEEEoverridecommandlockouts
\usepackage{cite}
\usepackage{amsmath,amssymb,amsfonts}
\usepackage{algorithmic}
\usepackage{graphicx}
\usepackage{subcaption}
\usepackage{textcomp}
\usepackage{comment}
\usepackage{balance}

\usepackage[whole]{bxcjkjatype}

\usepackage{xcolor}
\def\BibTeX{{\rm B\kern-.05em{\sc i\kern-.025em b}\kern-.08em
    T\kern-.1667em\lower.7ex\hbox{E}\kern-.125emX}}
\begin{document}

\title{Weather Synchronization in Digital Twin Environments for Shared VR Experience Using Commercial Metaverse Platforms}

\makeatletter
\newcommand{\linebreakand}{%
  \end{@IEEEauthorhalign}
  \hfill\mbox{}\par
  \mbox{}\hfill\begin{@IEEEauthorhalign}
}
\makeatother

\author{
\IEEEauthorblockN{Masanori Ibara}
\IEEEauthorblockA{
\textit{Cluster, Inc.}\\
Tokyo, Japan \\
m.ibara@cluster.mu
}
\and
\IEEEauthorblockN{Yuichi Hiroi}
\IEEEauthorblockA{
\textit{Cluster Metaverse Lab}\\
Tokyo, Japan \\
y.hiroi@cluster.mu}
\and
\IEEEauthorblockN{Takushi Kamegai}
\IEEEauthorblockA{
\textit{Cluster, Inc.}\\
Tokyo, Japan \\
t.kamegai@cluster.mu
}
\and
\IEEEauthorblockN{Yusuke Masubuchi}
\IEEEauthorblockA{
\textit{Kajima Corporation}\\
Tokyo, Japan \\
y-masubuchi@kajima.com
}
\linebreakand
\IEEEauthorblockN{Kazuki Matsutani}
\IEEEauthorblockA{
\textit{Kajima Corporation}\\
Tokyo, Japan \\
k.matsutani@kajima.com.sg
}
\and
\IEEEauthorblockN{Megumi Zaizen}
\IEEEauthorblockA{
\textit{Kajima Corporation}\\
Tokyo, Japan \\
zai@kajima.com
}
\and
\IEEEauthorblockN{Junya Morita}
\IEEEauthorblockA{
\textit{Kajima Corporation}\\
Tokyo, Japan \\
j.morita@kajima.com.sg
}
\and
\IEEEauthorblockN{Takefumi Hiraki}
\IEEEauthorblockA{
\textit{University of Tsukuba} \\
Ibaraki, Japan \\
\textit{Cluster Metaverse Lab}\\
Tokyo, Japan \\
hiraki@slis.tsukuba.ac.jp
}
}

\maketitle

\begin{abstract}
Digital twin technology creates bidirectional synchronization between physical and virtual environments, yet current implementations fail to provide authentic environmental experiences that enhance user presence in shared virtual spaces. While digital twin environments using commercial metaverse platforms for IoT sensor data visualization have been proposed, translating environmental information into meaningful sensory experiences remains largely unexplored, particularly lacking approaches for weather conditions that significantly influence spatial perception.
We developed a weather synchronization system that integrates real-time environmental data from ``The GEAR'' smart building with the Cluster metaverse platform, enabling shared VR experiences with authentic atmospheric immersion. Our system processes temperature, humidity, precipitation, wind speed, and solar radiation measurements to generate corresponding virtual environmental effects including dynamic sky rendering, precipitation particles, and ambient audio modulation. Performance evaluation demonstrated practical response times of 0.8-1.0 seconds for weather data transmission and reflection in the virtual environment.
This work establishes a foundation for seamless physical-digital space integration, potentially enhancing remote collaboration efficiency and facilitating more dynamic discussions in shared virtual workspaces.
\end{abstract}


\section{Introduction}
\begin{figure*}[t]
    \centering
    \includegraphics[width=\linewidth]{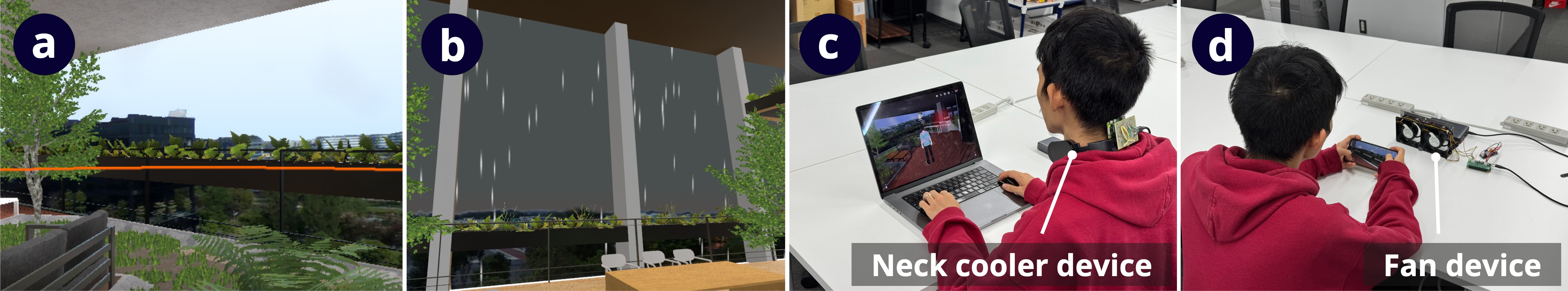}
    \caption{Weather-synchronized digital twin system integrating The GEAR smart building with Cluster metaverse platform. (a) Real-time weather visualization displaying atmospheric conditions from meteorological sensors within the shared VR environment. (b) Dynamic precipitation rendering with particle effects corresponding to actual rainfall intensity detected by sensors. (c,d) Enhanced multisensory VR experience through MetaGadget framework: an IoT neck cooler device provides thermal feedback reflecting temperature and solar radiation (c), while an IoT fan device simulates airflow matching actual wind speed measurements (d).}
    \label{fig:teaser}
\end{figure*}

A digital twin is a technology that creates a digital replica of physical objects or processes, characterized by bi-directional synchronization, where changes in one space are automatically reflected in the other~\cite{Kritzinger2018-kj,Grieves2014-dt,Fuller2020-dt}. Advances in commercial metaverse platforms such as VRChat~\cite{VRChat}, Resonite~\cite{Resonite}, and Cluster~\cite{Cluster} have established foundations for multi-user real-time interaction capabilities and shared virtual experiences. Recent research has achieved real-time data collection and visualization of environmental parameters within virtual spaces through the integration of IoT sensor networks with these platforms~\cite{Masubuchi2025}.

However, while digital twin environments using commercial metaverse platforms can effectively monitor and display sensor data, translating environmental information into meaningful sensory experiences remains largely unexplored. Weather conditions, in particular, constitute one of the most fundamental aspects of environmental experience, directly influencing human comfort, behavior, and spatial perception. Although weather visualization in VR/AR environments has been demonstrated~\cite{Ziegeler2001-metevr,Quick2020-meteovis,Heinrich2008-arweather}, existing approaches lack methods for translating environmental data into virtual weather phenomena that enable authentic environmental sharing between physically present and remote users in commercial metaverse-based digital twin environments. This represents a critical gap between data visualization and experiential immersion, presenting an opportunity to bridge this limitation through weather synchronization approaches.

Building upon previous work on sensor-metaverse integration, this paper addresses the experiential limitations of existing digital twin systems by developing a system that enables more natural sharing of experiences through weather synchronization (Fig.~\ref{fig:teaser}a, b). While our previous research~\cite{Masubuchi2025} established the technical foundation for integrating sensor networks with commercial metaverse platforms and visualizing environmental data, the current work extends this foundation by constructing a system that monitors weather conditions including temperature, humidity, and precipitation through a distributed sensor network. The system reflects these conditions within a virtual environment built on Cluster, a multi-device compatible metaverse platform accessible through smartphones, PCs, and VR-HMDs.

To achieve this integration, we developed a weather synchronization system that converts physical weather measurements into corresponding environmental representations in the virtual environment, including dynamic sky rendering, precipitation effects, and ambient audio modulation. In our evaluation, we measured the synchronization performance between weather data acquisition from sensors and its reflection in the virtual environment, achieving practical response times within the sensor update interval. Additionally, we implemented shared VR experiences with enhanced presence using the MetaGadget~\cite{Kurai2024-mr} framework. These results demonstrate the feasibility of creating an environment where remote users can share authentic real-time weather experiences of monitored physical spaces.

\section{Method}
\begin{figure}[t]
  \centering
  \includegraphics[width=\linewidth]{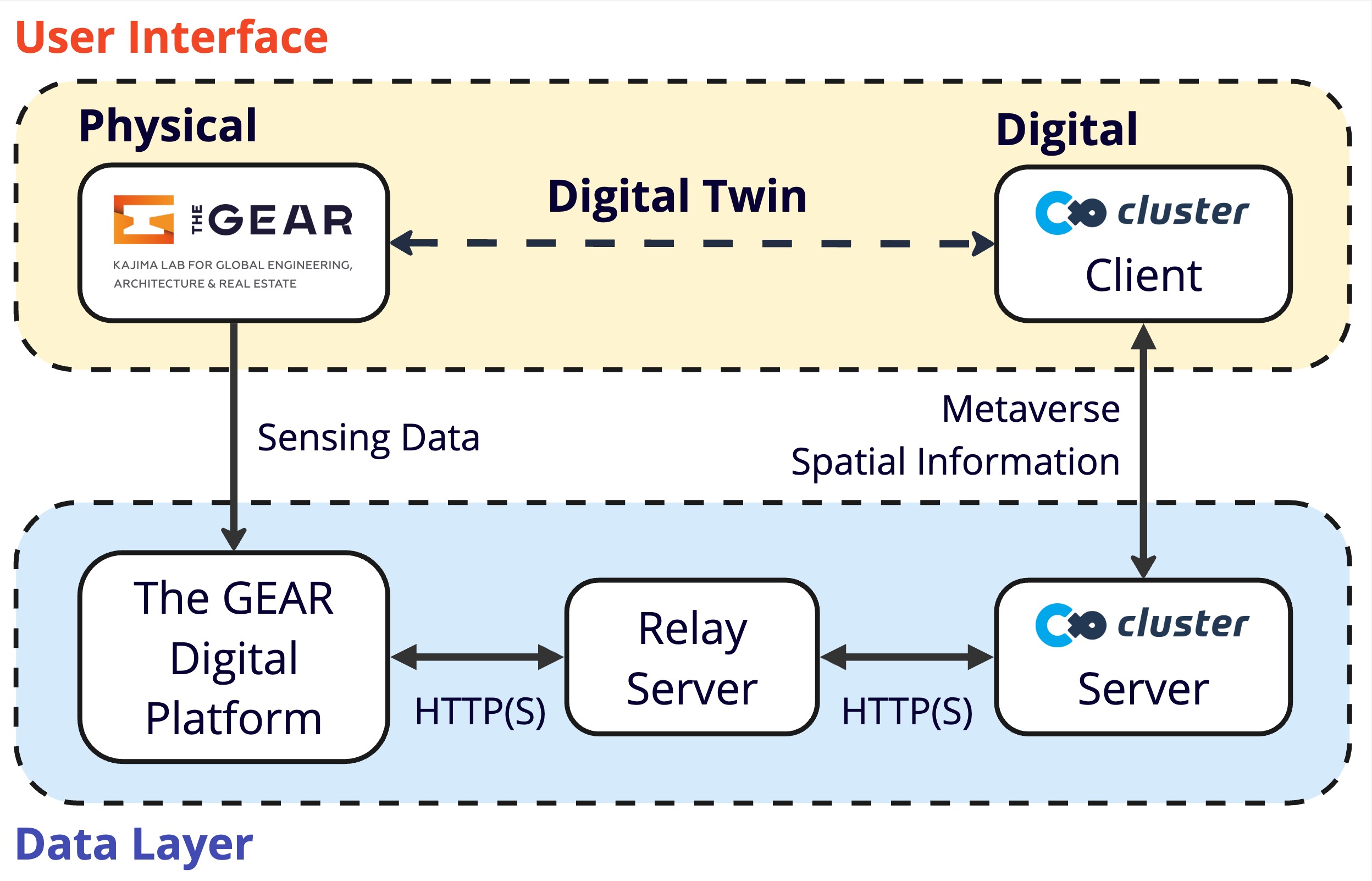}
  \caption{System architecture of our digital twin environment proposed in~\cite{Masubuchi2025}. Sensor data from The GEAR smart building is collected by The GEAR Digital Platform (TGDPF), while spatial information in the metaverse is exchanged between the Cluster client and server. These two systems are integrated through the relay server that bridges TGDPF and the Cluster server infrastructure.}
  \label{fig:architecture}
\end{figure}

In this section, we describe the implementation of our weather-synchronized digital twin system that integrates real-time environmental conditions with virtual space rendering. A diagram of the system is shown in Fig.~\ref{fig:architecture}.

Our implementation builds upon the established system components from our prior work~\cite{Masubuchi2025}, utilizing The GEAR~\cite{The_GEAR} as our smart building testbed, The GEAR Digital Platform (TGDPF) for IoT sensor data collection, and Cluster as our metaverse platform. The existing relay server that facilitates communication between TGDPF and Cluster is maintained to ensure system modularity and interoperability. Since external communication functionality in Cluster is conducted through the Cluster server rather than directly from individual user client applications, the actual communication between client applications and TGDPF involves routing through two servers: the Cluster server and the relay server.

We implemented weather data acquisition by accessing Weather Station sensors connected to TGDPF. In addition to the standard temperature and humidity measurements utilized in previous work, we extended the data collection to include Rain (precipitation volume), Wind (wind speed), and IR (solar radiation energy) measurements. These sensors provide continuous monitoring of atmospheric conditions at five-minute intervals, with the acquired data transmitted through the existing relay server infrastructure to the Cluster virtual environment.

\begin{figure}[t]
  \centering
  \includegraphics[width=\linewidth]{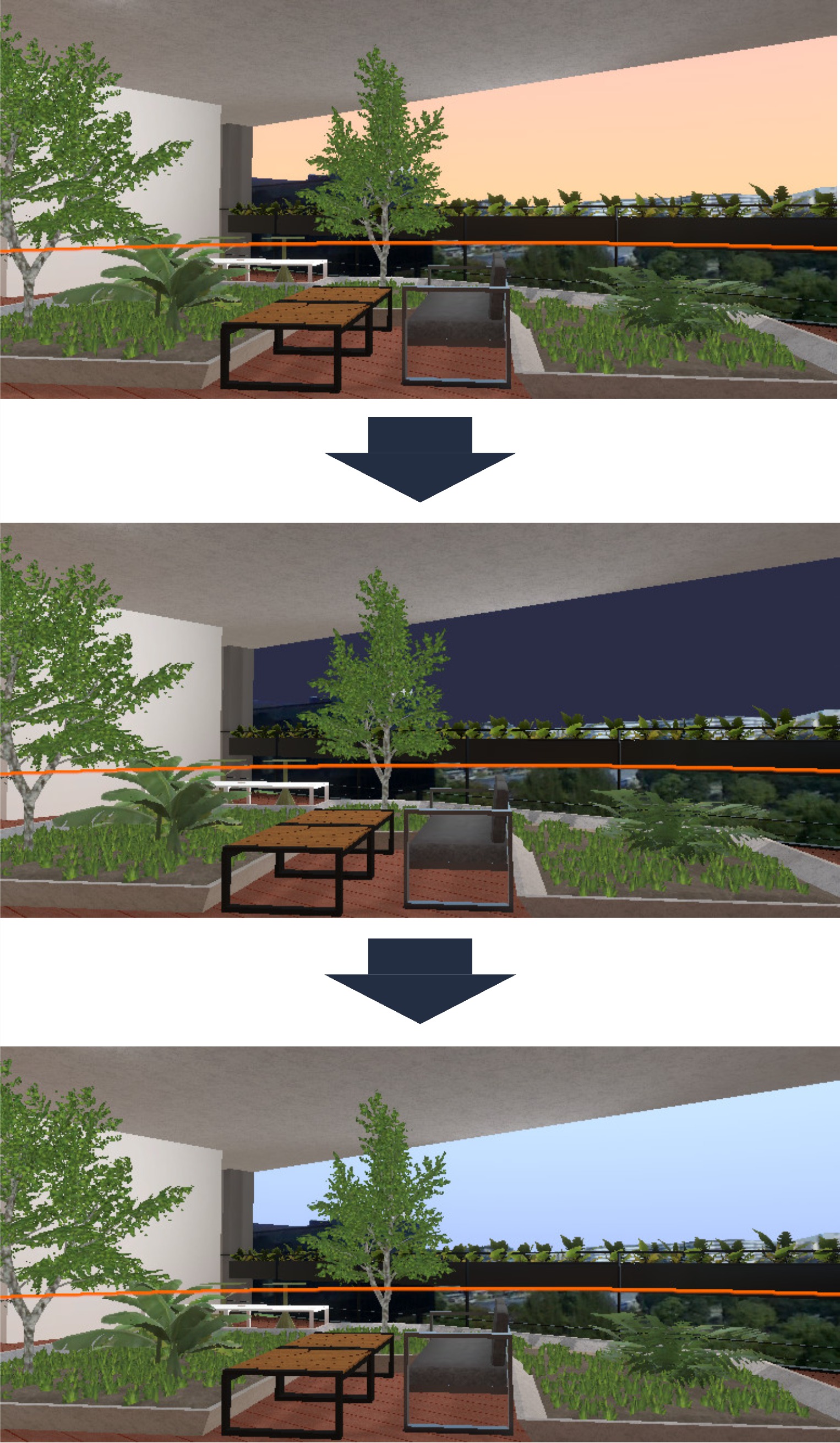}
  \caption{Temporal weather visualization changes in the virtual environment showing the view from inside The GEAR building. The virtual weather conditions dynamically evolve based on real-time sensor data and temporal progression: (top) evening scene with warm atmospheric lighting, (middle) nighttime scene with reduced visibility and artificial lighting, (bottom) morning scene with early daylight conditions.}
  \label{fig:weather_change}
\end{figure}

The acquired weather data undergoes specific processing to generate corresponding environmental effects within the virtual space. Solar radiation energy measurements are processed to determine cloud coverage conditions, enabling dynamic adjustment of sky rendering based on actual sunlight intensity. Precipitation data is processed through a mapping function that converts rainfall measurements ranging from 0 to 100 mm/h to normalized particle system parameters with float values from 0 to 1. Rainfall values exceeding 100 mm/h are capped at the maximum particle density to prevent system overload. Wind speed measurements are integrated with the audio processing system established in prior work, dynamically modulating wind sound effects to correspond to actual wind conditions. In addition to the weather sensor data, we obtain current date and time information from the system, which is utilized to calculate solar elevation and azimuth angles, with appropriate timezone corrections applied to accurately reflect the sun's position in the virtual environment.

The processed weather parameters are applied to the virtual environment through a real-time environmental simulation system implemented within Cluster. This system dynamically adjusts multiple environmental aspects including sky conditions based on solar radiation measurements, precipitation particle effects corresponding to rainfall intensity, solar positioning reflecting actual time and date, and ambient audio effects modulated by wind speed data. The synchronization ensures that users in the virtual environment experience weather conditions that accurately represent the real-time atmospheric state of The GEAR building. Figure~\ref{fig:weather_change} illustrates the temporal changes in weather conditions within the virtual space, demonstrating how the environmental visualization evolves over time in response to sensor data from the building.

\section{Experiment}
\begin{table}[t]
    \centering
    \caption{Response Time for Weather Sensor Data Retrieval Requests.}
    \scalebox{1.1}[1.1]{
    \begin{tabular}{|l|r|}\hline
     Measurement Type & Response Time [ms]\\ \hline
    IR & 1,039.43 $\pm$ 2.79  \\ 
    Rain & 892.97 $\pm$ 1.57  \\ 
    Wind & 840.88 $\pm$ 2.14 \\ \hline
    \end{tabular}
    }
    \label{tab:latency}
\end{table}

To evaluate the effectiveness of environmental information visualization in our proposed system, we measured the data transmission latency between the Cluster client and The GEAR Digital Platform (TGDPF). We conducted 100 information retrieval requests for each latency measurement from the Cluster client to TGDPF and measured the response times for IR, Rain, and Wind sensors under identical conditions using weather sensor devices.

The weather sensor device employed in our system provides comprehensive meteorological data including infrared radiation (IR), precipitation (Rain), and wind speed (Wind) measurements. This weather sensor unit is installed on the rooftop to monitor these three distinct environmental parameters.

The mean response times and standard deviations for each measurement type are presented in Table~\ref{tab:latency}. Our measurements revealed mean response times of 1,039.43 ms (SD = 2.79 ms) for IR data retrieval, 892.97 ms (SD = 1.57 ms) for Rain data, and 840.88 ms (SD = 2.14 ms) for Wind measurements.

Direct data retrieval from TGDPF, bypassing the Cluster server and relay server, yielded response times of 212.11 ms (SD = 3.84 ms) for the weather sensor as a whole. This difference reflects the architectural characteristics of each data flow path: the normal flow involves Cluster client $\rightarrow$ TGDPF $\rightarrow$ IoT sensor device gateway $\rightarrow$ TGDPF $\rightarrow$ Cluster client, while direct retrieval follows a path of Local $\rightarrow$ IoT sensor device gateway $\rightarrow$ Local. The variation in response times among different measurement types (IR, Rain, and Wind) can be attributed to the varying computational complexities required for processing each environmental parameter within the sensor device gateway.

Based on these measurements, our implemented data relay server demonstrates practical response times for an environmental information visualization system, achieving latencies of approximately 0.8-1.0 seconds for all weather-related measurements (IR, Rain, and Wind). These response times are adequate for weather data visualization applications, providing sufficient real-time capability for weather information reflection. Additionally, the latency through the relay server could potentially be improved by optimizing server deployment regions and other infrastructure configurations. However, it should be noted that the weather sensor updates its data only every 5 minutes, which constitutes a constraint on the overall system responsiveness. This limitation is attributed to the smart building sensor configuration, and our system is capable of supporting more frequent weather information updates.

\section{Application}
We developed a system that feeds weather sensor information from ``The GEAR'' into Cluster, a metaverse platform, enabling real-time modification of background rendering in the virtual space to synchronize with the real-world environment. Using this system, we explored application scenarios that provide users with VR experiences of shared virtual spaces.
In this study, we targeted ``K/PARK,'' the semi-outdoor workspace of The GEAR, and implemented two experiences that provide users with immersion through IoT device control integrated with visually and auditorily represented weather information. MetaGadget~\cite{Kurai2024-mr} framework was utilized for IoT device control.

The first application enables users to experience local weather conditions through temperature presentation to the body (Fig.~\ref{fig:teaser}c). This application employs the neck cooler device proposed in MetaGadget~\cite{Kurai2024-mr} and controls the thermal sensation presented to the user's neck based on the temperature and solar radiation conditions within the physical office space of K/PARK. Since K/PARK is located in Singapore, which has a tropical climate, users in remote locations can realistically experience this climate.

The second application allows users to experience local weather conditions through wind presentation (Fig.~\ref{fig:teaser}d). This application utilizes the fan device proposed in MetaGadget~\cite{Kurai2024-mr} and controls the airflow presented to users according to the wind conditions within the K/PARK space. K/PARK realizes an open workspace where wind flows through by combining natural ventilation with ceiling fans, enabling users in remote locations to realistically feel these characteristics.

\section{Conclusion}
In this paper, we developed a weather-synchronized digital twin system that enables shared VR experiences with authentic environmental immersion by reflecting real-time weather conditions from ``The GEAR'' smart building into the Cluster metaverse platform. Our system successfully bridges the gap between physical environmental data and virtual experiential representation through comprehensive weather parameter integration including temperature, humidity, precipitation, wind speed, and solar radiation measurements.

The implementation demonstrates practical feasibility with weather sensor data being transmitted and reflected in the metaverse environment through our data relay server within 0.8--1.0 seconds, providing adequate real-time capability for weather synchronization applications. The acquired weather data is processed and mapped to corresponding environmental effects in the virtual space, including dynamic sky rendering based on solar radiation intensity, precipitation particle effects corresponding to rainfall measurements, solar positioning reflecting actual time and date, and ambient audio effects modulated by wind speed data.

Furthermore, we developed application scenarios utilizing the MetaGadget~\cite{Kurai2024-mr} framework to enhance user immersion through IoT device integration, including neck cooler devices for temperature experience and fan devices for wind sensation. These applications demonstrate the potential for creating authentic environmental experiences that allow remote users to genuinely feel the atmospheric conditions of monitored physical spaces.

Future work will investigate the extent to which our developed VR experience contributes to users' perception of sharing the same spatial atmosphere between smart building occupants and remote users in the metaverse environment. Through such evaluation, we anticipate establishing a foundation for new value creation through seamless integration of physical and digital spaces, potentially enhancing remote collaboration efficiency and facilitating more dynamic discussions in shared virtual workspaces.

\section*{Acknowledgment}
This work was partially supported by JST Moonshot Research \& Development Program Grant Number JPMJMS2013 and JST ASPIRE Grant Number JPMJAP2327, Japan.

\balance
\bibliographystyle{abbrv}

\bibliography{ibara_ismar2025}
\end{document}